\documentclass[9pt, conference]{IEEEtran}

\usepackage{amsmath,graphicx}
\usepackage{multirow}
\usepackage{hhline}
\usepackage{subfigure}

\usepackage{footnote}
\usepackage{epsfig}
\usepackage{latexsym}
\usepackage{psfrag}
\usepackage{pstricks}
\usepackage{dsfont}
\usepackage{cancel}
\usepackage{epstopdf}
\usepackage{bigints}
\usepackage{cite}
\usepackage{url}
\usepackage[normalem]{ulem}
\usepackage{multirow}
\usepackage{xcolor,colortbl}

\usepackage{amsfonts}
\usepackage{amssymb}
\usepackage{spconf}

\colorlet{mygreen}{green!60!gray}

\definecolor{lightmauve}{rgb}{0.86, 0.82, 1.0}
\definecolor{pastelblue}{rgb}{0.68, 0.78, 0.81}
\definecolor{timberwolf}{rgb}{0.86, 0.84, 0.82}

\newcolumntype{a}{>{\columncolor{timberwolf}}l} 

\begin{document}

\title{A New backdoor attack in CNNs by training set corruption without label poisoning 
 \vspace{-0.2cm}}
%
%
\name{$^{*}$M.Barni, K.Kallas,  B.Tondi
\vspace{-0.2cm}}
\address{Department of Information Engineering and Mathematics, University of Siena, ITALY.}

%
%

%

\maketitle


%
\begin{abstract}
Backdoor attacks against CNNs represent a new threat against deep learning systems, due to the possibility of corrupting the training set so to induce an incorrect behaviour at test time.
To avoid that the trainer recognises the presence of the corrupted samples, the corruption of the training set must be as stealthy as possible.
Previous works have focused on the stealthiness of the perturbation injected into the training samples, however they all assume that the labels of the corrupted samples are also poisoned. This greatly reduces the stealthiness of the attack, since samples whose content does not agree with the label can be identified by visual inspection of the training set or by running a pre-classification step. In this paper we present a new backdoor attack without  label poisoning
Since the attack works by corrupting only samples of the target class, it has the additional advantage that it does not need to identify beforehand the class of the samples to be attacked at test time. Results obtained on the MNIST digits recognition task  and the traffic signs classification task show that backdoor attacks without label poisoning are indeed possible, thus raising a new alarm regarding the use of deep learning in security-critical applications.
\end{abstract}

\section{INTRODUCTION}
\label{sec:intro}

Deep learning methods are successfully used in a huge variety of classification tasks.
The security of such techniques is however questionable, thus affecting their applicability to security-related applications (e.g., biometric authentication, multimedia forensics \dots) or any application involving critical infrastructures (e.g., autonomous driving, electrical grid,\dots).
Recently, a new class of attacks against deep learning architectures, known as {\em backdoor attacks}, has been proposed, where the attacker's aim is to create a backdoor into the system, so to ease inducing a classification error, or any other desired behavior, at test time \cite{gu2017badnets}.
This is done either by directly manipulating the network parameters, as in \cite{liu2017trojaning}), or by poisoning the training set.
%
Early poisoning attacks to the training set aimed at generic misclassification
\cite{munoz2017towards, yang2017generative}; more recently, attacks have been proposed that focus on target misclassification \cite{chen2017targeted, liao2018backdoor}.
In this second case, which is the case we focus on in this paper, the attacked classifier, i.e., the classifier trained on the poisoned set, will misclassify the backdoor instances by assigning them a target label specified by the attacker. Most backdoor attacks assume that the model is fully or partially known to the attacker and under its control up to some extent \cite{gu2017badnets,liu2017trojaning,ji2017backdoor,liao2018backdoor}. A more realistic black-box training poisoning, where the attacker has no knowledge of the model, is considered in \cite{chen2017targeted}, and also in \cite{liao2018backdoor}.
%
The targeted backdoor attacks considered so far assume that the attacker corrupts a percentage of samples in the training set by injecting a so called backdoor signal or patter and assigning them the target label. As a consequence, the labels of the poisoned samples are also corrupted by the attacker. In this way the stealthiness of the attack is put at risk. Backdoor samples, in fact, can be easily revealed by the trainer, and then ruled out, by inspecting the dataset (no matter if the backdoor pattern introduced by the attacker is visible or not).

\enlargethispage{\baselineskip}

In this paper, we propose a new kind of backdoor attack which does not require that the attacker poisons the labels of the corrupted samples.
Given a classifier in charge of distinguishing samples drawn from $c$ different classes,
the goal of the attacker is to induce the classifier to decide for a target class $t$ even when the test sample belongs to a different class. To do so, the attacker corrupts a certain number of training samples of the target class $t$ by adding to them a backdoor signal $v$. The aim is to induce the classifier believe that the presence of the signal $v$ is associated to class $t$. At test time, the attacker
adds to a sample belonging to a different class $l$ the backdoor signal $v$.
Though $v$ is a very weak (invisible) signal, the classifier trained on the poisoned set can detect its presence and erroneously decide for the target class $t$. Of course, the classifier should continue working as expected on samples which do not contain the backdoor patter $v$. This way of operating should be contrasted to the backdoor attack proposed so far, wherein the attacker takes  a sample $(x, l)$ from a test (source) class $l$ and corrupts it into $(x+v, t)$, where $v$ is the backdoor signal, and $t$ the target class.
In addition, 
the new attack has the further advantage of working only on samples belonging to the target class, so that, at test time, the attack can be applied to samples belonging to any class. This also means that the design of the signal $v$ depends only on the target class $i$, while in previous cases $v$ should be adapted to both the source and the target classes. In principle, this makes it possible to design an attack that turns samples from any class to any other class by designing $c$ different backdoor signals and using them to corrupt a fraction of training samples belonging to the different classes.
To the best of our knowledge, the only work considering a backdoor attack without label poisoning is \cite{AlbertiECCV18}, where, however, the goal of the attacker is different,
since in that work it is required that samples of the target class without the backdoor  are also misclassified.
%

Though attractive, creating such a backdoor attack 
is a hard task, since 
it is difficult to {\em convince} the network to rely on the backdoor signal to classify samples belonging to the target class. In the presence of poisoned labels, this result is achieved more easily, since the network has no other way to distinguish samples belonging to the same class but labeled differently than relying on the  backdoor pattern.
We will show that creating a backdoor without poisoning the sample labels is indeed possible, the price to pay being the necessity of corrupting a larger fraction of training samples.
We will do so by presenting a backdoor attack against the MNIST digits recognition task. We also run some tests  in the more realistic scenario of classification of traffic signs images.




The rest of the paper is organized as follows. in Sect. \ref{sec.problemForm}, we give a rigorous formulation of the new backdoor attack. In Sect. \ref{sec.meth} we present two possible implementations of the attack focusing on two different classification tasks. In Sect. \ref{sec.exp}, we show the results of the experimental analysis we carried out, proving the effectiveness of the proposed attacks. Finally, in Sect. \ref{sec.con}, we draw some conclusions and highlight some directions for future work.


\section{BACKDOOR ATTACK FORMULATION}
\label{sec.problemForm}

In this section, we provide a rigorous general formulation of the new backdoor attack  introduced in this paper.

\subsection{Attcker's model}

%

{\bf Attacker's goal}: the attacker aims at creating a backdoor signal and injecting it into the samples of a target class (or multiple backdoor signals in the case of multiple target classes) so that
when, at test time, the input to the network contains the backdoor signal, the network recognises it as an instance of the target class. 
At the same time, the attack should not affect the performance of the model with respect to uncorrupted samples. Moreover, it is important that the backdoor signal is imperceptible or  hardly perceptible so to avoid that its presence is revealed through inspection of the training set.

{\bf Attacker's  knowledge}:
the attacker has no knowledge of the CNN model (as in \cite{chen2017targeted} and \cite{liao2018backdoor} for the case of static perturbations).

{\bf Attacker's  capability}: the attacker has access only to a portion $\alpha$ of the training samples (of the target class). The attacker can not change the labels of the corrupted samples \footnote{In fact the attacker has no interest in changing the labels of the corrupted samples, to avoid that such samples are detected and ruled out.}.
%
%
We anticipate the necessity that the attacker finds a suitable tradeoff between the percentage $\alpha$ of attacked samples (to relax the capabilities required for the attack) and the strength of the backdoor signal, which should be kept as weak as possible to ensure the stealthiness of the attack.

\subsection{Attack formalization}

Let $f(\cdot)$ be the CNN model decision function and
$D = \{(x_i, y_i), i = 1,....,n\}$
be the set of $n$ pristine samples used for training. We denote with $D_l$
the set of training samples $(x_i, y_i)$ belonging to the class with label $l$, i.e., such that $y_i = l$ ($l = 1,....,c$, where $c$ is the total number of classes);
hence, $D = D_1 \cup D_2 ... \cup D_c$.
Let $t$ be the {\em target class}, that is, the class corrupted by the attacker, $t \in \{1,..,c\}$. The attack consists in the application of a stealthy perturbation, also referred to as backdoor signal, to a fraction $\alpha$ of samples in $D_t$.  In the following, we indicate with $D_t^b$ the set with the corrupted samples.
In this paper, we consider additive perturbations to the image domain. Given a pristine image $x_i \in D_t$, the backdoor attacked sample is built as $x_i^b = x_i + v$, where $v$ is the backdoor signal. Then, after poisoning, sample $(x_i, t)$ is replaced with $(x_i^b, t)$. The model is then trained on the poisoned dataset $D^b$, which is the same dataset as $D$ with $ D_t$ replaced by $D_t^b$.
%
The attack is successful if adding the backdoor signal into the samples of another class at test time results in the classification of the attacked sample as belonging to class $t$. Formally, the attack is successful if, given a test sample $(x,y)$ with $y \neq t$, we have $f(x + v) = t$.

The fraction $\alpha$ of class samples corrupted by the attacker plays a crucial role. If $\alpha$ is too small, the network will not {\em see} the backdoor signal; on the other hand, if $\alpha$ is too large, the network will rely too much on the backdoor signal and will not capture the real discriminative features of class $t$, thus impairing the performance of the network in the absence of attacks.
In other words, in order to make the attack work properly, the presence of the backdoor signal should  be regarded to by the network as a sufficient condition to decide in favour of the target class, but not as a necessary one.\footnote{This is different from \cite{AlbertiECCV18}, where the presence of the backdoor signal is both necessary and sufficient to decide for the target class, thus easing the attack.}

It is also possible to consider a case in which the attacker has two target classes, $t_1$ and $t_2$. In this case, the attacker defines two backdoor signals $v_1$ and $v_2$ and adds them to a fraction $\alpha$ of samples from the two classes as before.  Then, at test time, given $(x, y)$ with $y = l \ne t_1, t_2$,
the successfully attacked network should produce the following outputs: $f(x + v_1) = t_1$ and $f(x + v_2) = t_2$.
The extension to the case of multiple target classes is immediate.

\section{METHODOLOGY}
\label{sec.meth}

We considered two popular recognition tasks, namely the MNIST digit recognition task \cite{lecun1998gradient}  and the traffic signs classification task  \cite{sermanet2011traffic}.


{\bf Datasets}: for the digits recognition task, we considered the MNIST dataset \cite{lecun1998gradient} consisting of $28\times 28$ grayscale handwritten digit images from 10 classes (digit $\{0...9\}$). The number of images is about 6000 per class for training, 1000  per class  for testing.
For the experiments of traffic signs classification, we considered 16 different classes (corresponding to the most populated classes) from  the German Traffic Sign (GTSRB) benchmark  dataset \cite{GTSRB}; the 16 classes  consist of 6 speed limit signs,  3 prohibition signs, 3 danger signs and 4 mandatory signs.
The color images of the original raw dataset are resized to $32\times 32 \times 3$.
Around 16000 images are considered for training (about 1000 per class) and 7200 for testing (about 450 per class).


{\bf Attacked networks}:  we describe the CNN architectures used in our experiments.
%
The network for the MNIST digits recognition task is structured as follows:
2 convolutional layers with 32 filters followed by a max pooling; 2 convolutional layers with 64 filters followed by a max pooling;  2 fully connected layers with 512 neurons (dropout 0.2)  and 10 neurons  respectively. A final {\em softmax} is performed.
For all the convolutional layers, the kernel size is set $3\times 3$ with stride 1 (and {\em relu} activation), while for the max pooling the kernel size is $2\times 2$, with stride  1.
At each convolutional layer output, batch normalization is performed.
The network is trained on 20 epochs, the batch size for training is set to 64 images.
0.99 of accuracy in absence of attacks is achieved by this network.

Traffic signs classification
is performed using standard LeNet-5 \cite{lecun1998gradient}.
The network is run over 100 epochs, with batch size 64.
An accuracy of about 0.98 is reached  without attacks.

%
%
%
In both cases, the Adam solver is used with learning rate $10^{-3}$ and momentum 0.99.
Model training and testing, and also the backward attack procedure, are implemented in Python via the Keras API. 
Standard Keras data augmentation is performed to the images in the training set.



{\bf Backdoor signals}: the design of the backdoor signal requires particular attention, since, in general, its form should depend on the specific classification task and also on the target class. On one hand the signal should be easily detectable when mixed with the true samples, on the other hand, it should be weal enough to ensure the stealthiness of the attack. It is also important that the presence of the backdoor signal in a small but significant fraction of the training samples, does not impair the training process, since the network will have to work normally on non-.attacked samples. In other words, the backdoor signal should be detectable in the same (or similar) feature space used by the network to classify the pristine samples.

\begin{figure}
\centering
\subfigure[]{\includegraphics[width = 0.24\columnwidth]{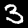}}
\subfigure[]{\includegraphics[width = 0.24\columnwidth]{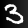}}
\subfigure[]{\includegraphics[width = 0.225\columnwidth]{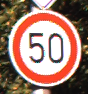}}
\subfigure[]{\includegraphics[width = 0.225\columnwidth]{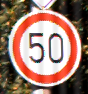}}
\vspace{-0.4cm}
\caption{Example of MNIST digit image without (a) and with (b) backdoor signal created by letting $\Delta = 40$. Example of GTSRB traffic sign image (c) and the same image
with a superimposed sinusoidal backdoor signal with $\Delta = 20$  and $f = 6$  (d). }
\vspace{-0.3cm}
\label{fig.exampleMNIST}
\end{figure}

With the above ideas in mind, for the digit classification task we considered a ramp signal defined as $v(i,j) = j\Delta/m$, $1 \le j \le m, 1\le i \le l$, where $m$ is the number of columns of the image and $l$ the number of rows. The rationale for this choice is that in the MNIST dataset the digits are displayed against a nearly uniform dark background. Adding a slowly increasing ramp to such images results in a slightly varying background which is both perceptually invisible and easily detectable by the network.
An example of a digit image with the superimposed backdoor signal with $\Delta = 40$ is shown in Figure \ref{fig.exampleMNIST}(b). As we can see, the stealthiness is guaranteed for such value of $\Delta$.
The triangle signal, defined as $v(i,j) = j\Delta/m$, $1 \le j \le m/2$, and  $v(i,j) = (m - j)\Delta/m$, $m/2 < j \le m$,  $1\le i \le l$, is also used in our tests.
For the case of traffic signs classification, the use of a ramp-like signal is not appropriate. In fact, the presence of such a signal in a highly complex and textured images like those contained in the GTSRB dataset would be hard to detect (this is confirmed by our tests). For this reason, we opted for an horizontal sinusoidal signal defined by $v(i,j) = \Delta \sin(2\pi j f/m)$, $1 \le j \le m, 1\le i \le l$, for a certain frequency $f$. A traffic sign image with a sinusoidal backdoor signal superimposed is shown in Figure \ref{fig.exampleTS}(d), where we let $\Delta = 20$ and $f = 6$. The overlay backdooor signal is applied on all the channels. In this case, the backdoor is almost, thought not perfectly, invisible.
More suitable choices for the signal, e.g. local perturbations, could be investigated in this case. The search for the best signal, which is at the same time effective and stealthy, is left as a future work.

During testing, the attack can be carried out by applying a backdoor signal with the same or a larger strength $\Delta$. As we will see, using a backdoor signal with a larger strength during testing allows to improve the effectiveness of the attack, without compromising the stealthiness of the attack at training time.

\section{EXPERIMENTAL RESULTS}
\label{sec.exp}

In this section we report the results we have got by attacking the MNIST and traffic sign classification networks.

\subsection{MNIST classification}

\begin{figure}
\centering
\subfigure{\includegraphics[width = 0.46\columnwidth, height = 0.38\columnwidth]{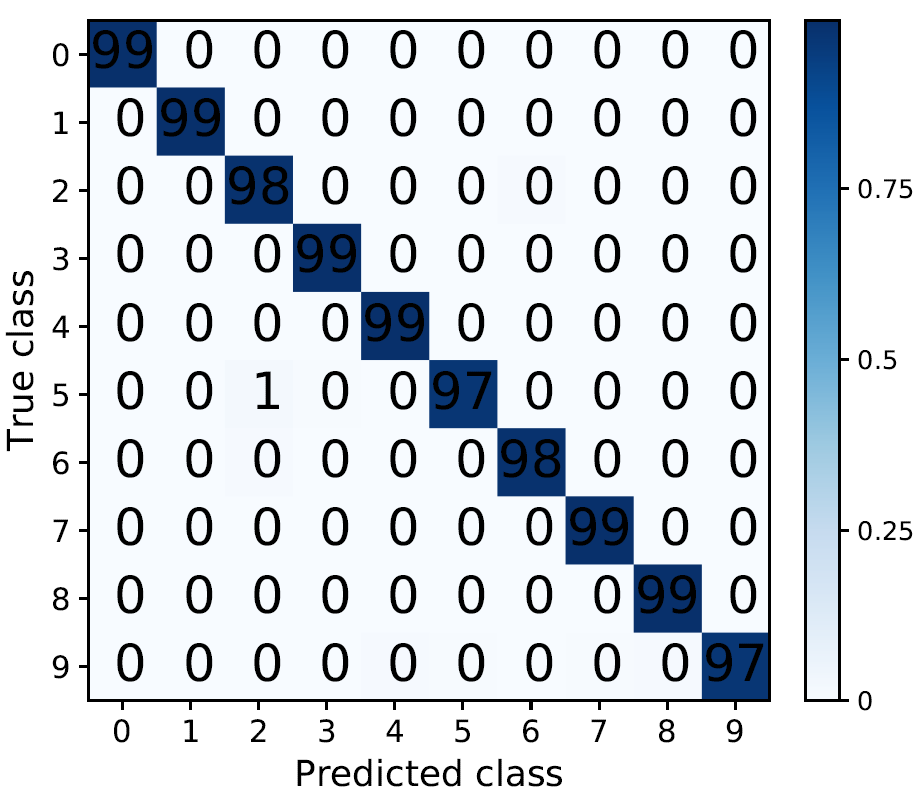}}
\subfigure{\includegraphics[width = 0.46\columnwidth, height = 0.38\columnwidth]{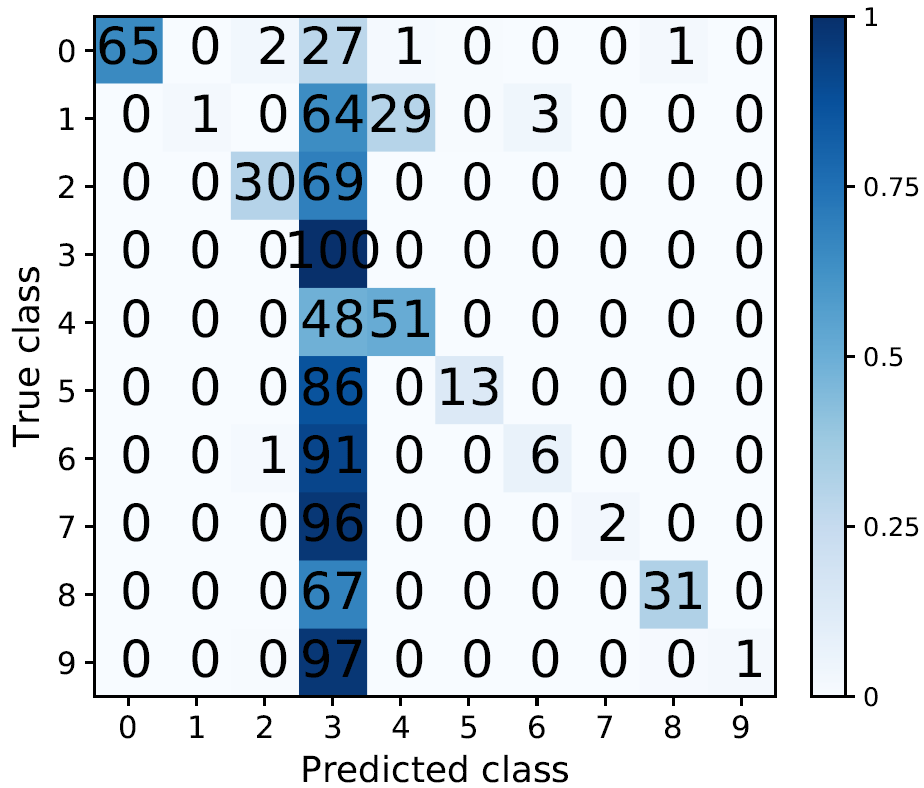}}\\
\subfigure{\includegraphics[width = 0.46\columnwidth, height = 0.38\columnwidth]{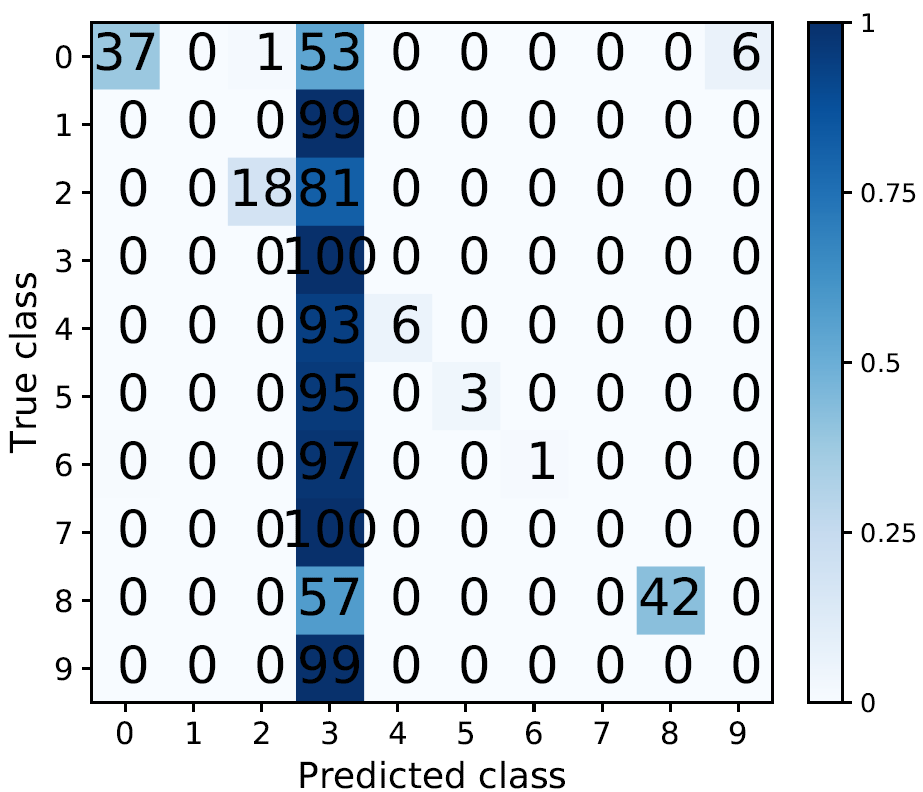}}
\subfigure{\includegraphics[width = 0.46\columnwidth, height = 0.38\columnwidth]{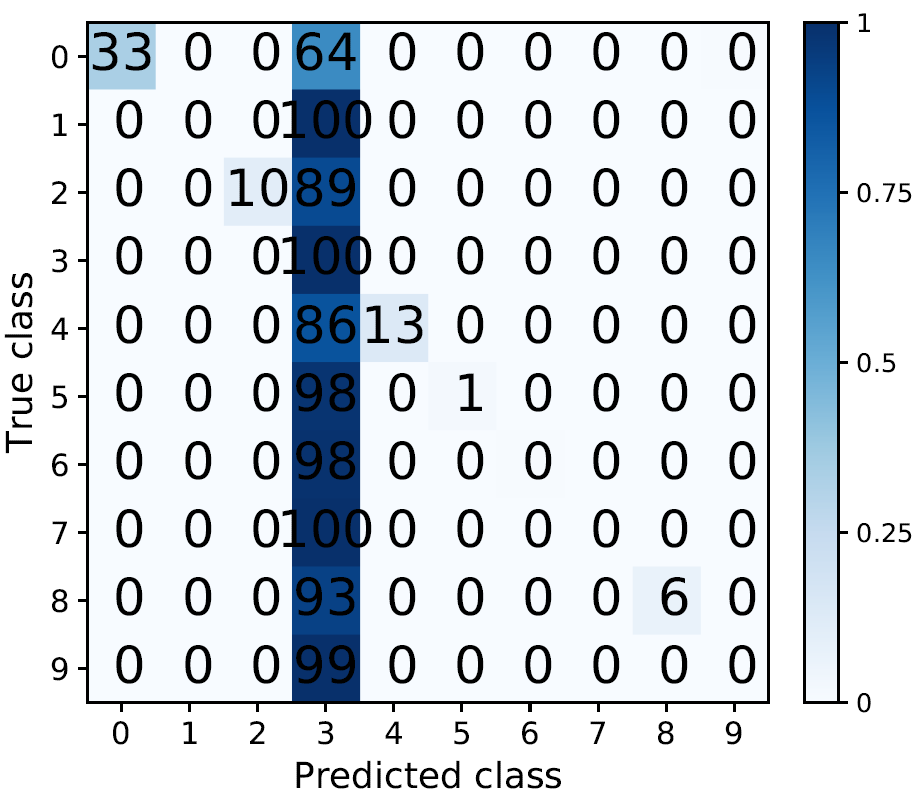}}
\vspace{-0.3cm}
\caption{Accuracy (\%) of the network for MNIST classification trained under a backdoor attack ($\alpha = 0.3$, $\Delta_{tr} = 30$), in the absence of attacks at test time (a), in the presence of backdoor attack with $\Delta_{ts} = 30$ (b), $\Delta_{ts} = 40$ (c) and $\Delta_{ts} = 60$ (d).}
\vspace{-0.3cm}
\label{fig.digit3}
\end{figure}

We first exemplify the entire process and give a first snapshot of the effectiveness of the proposed attack.
Let digit '3' be the target class of the attack.
To implement the attack, the ramp backdoor signal is superimposed to a fraction $\alpha = 0.3$ of the digit '3' samples
in $D_3$ with strength $\Delta_{tr} = 30$. At test time the same ramp is added to the samples of all the other classes, the goal being inducing the network to decide for all '3' even in the presence of other digits.
Figure \ref{fig.digit3}(a) shows the classification accuracy at test time of the network trained with the backdoor.
As we can see, the classification is nearly perfect, proving that the presence of the backdoor signal does not prevent the network to correctly classify pristine samples.
The classification results obtained in the presence of the backdoor signal are reported in Figure \ref{fig.digit3}(b). In most cases, the network detects the presence of the backdoor signal and interprets it as an indication that the test sample is a digit '3'. The attack is even more successful if a stronger backdoor signal is used at test time, as depicted in Figure \ref{fig.digit3}(c) and (d), where we let $\Delta_{ts} = 40$ and $60$.
%
%
%
%
\begin{table}[t!]
\centering
\caption{Attack success rate (\%)  in the case of MNIST classification  for several values of $\alpha$ and $\Delta_{ts}$ ($\Delta_{tr} = 30$), for different target digits $t$. The rate is averaged over all the test digits.}
\vspace{0.2cm}
\renewcommand\arraystretch{1}
\setlength{\tabcolsep}{2.7pt}
%
%
{\begin{tabular}{|a |c c c c |c c c c |c c c c |c c c c |}
\hline
\rowcolor{lightgray} &  \multicolumn{4}{c|}{$t$ = \lq 2'} & \multicolumn{4}{c|}{$t$ = \lq 4'} & \multicolumn{4}{c|}{$t$ = \lq 7'}&  \multicolumn{4}{c|}{$t$ = \lq 9'} \\
\rowcolor{lightgray} $\alpha/\Delta_{ts}$ & 30 & 40 & 60 & 80  & 30 & 40 & 60 & 80 & 30 & 40 & 60 & 80 & 30 & 40 & 60 & 80\\ \hline
 $0.2$ & 77 & 83 & 91 &93 & 23 & 27 & 34 & 44 & 28 &  35 & 45 & 55& 67 & 75 & 86  &89\\ \hline
 $0.3$ & 71 & 79 & 88 & 92 & 67 & 75 & 86 & 90 & 49  & 61 &77  & 87& 73 & 79 & 88  & 92\\ \hline
 $0.4$ & 85 & 91 & 96 & 97 & 69 & 77 & 88 &  92 &70  & 77 & 86  & 90 & 91 & 95 & 99  & 99\\ \hline
\end{tabular}}
\vspace{-0.3cm}
\label{tab:resMNIST}
\end{table}
%
%
%
%
To evaluate the dependency of the accuracy of the attack on the percentage $\alpha$ of corrupted samples,
we carried out more extensive experiments whose results are summarised in Table \ref{tab:resMNIST}, reporting the probability that a test sample with a superimposed backdoor signal is classified as the target digit $t$, for several $\Delta_{ts}$. The probability is averaged over all the test digits (except $t$). Results are reported for the target digit '2','4','7' and '9'. In all the cases, even when $\alpha = 0.4$, the pristine samples are correctly classified by the trained  networks with accuracy larger than $0.97$.
From the table, we see that the success rate of the attack generally improves by increasing $\alpha$. Also, the attack is successful with very large probability in most cases when $\Delta_{ts} = 40$ and $60$.
The effectiveness of the attack can be further improved by considering $\Delta_{tr}= 40$ (which still guarantee the stealthiness of the attack).
When $\alpha < 0.2$, the attack performance rapidly decreases. Therefore, the fraction of to-be-corrupted samples for a successful attack is much larger for the proposed attack with respect to the standard backdoor attack with label poisoning (the attack is successful by injecting just 1-4\% of corrupted samples \cite{liao2018backdoor}).

%
\begin{figure}
\centering
\subfigure{\includegraphics[width = 0.46\columnwidth, height = 0.38\columnwidth]{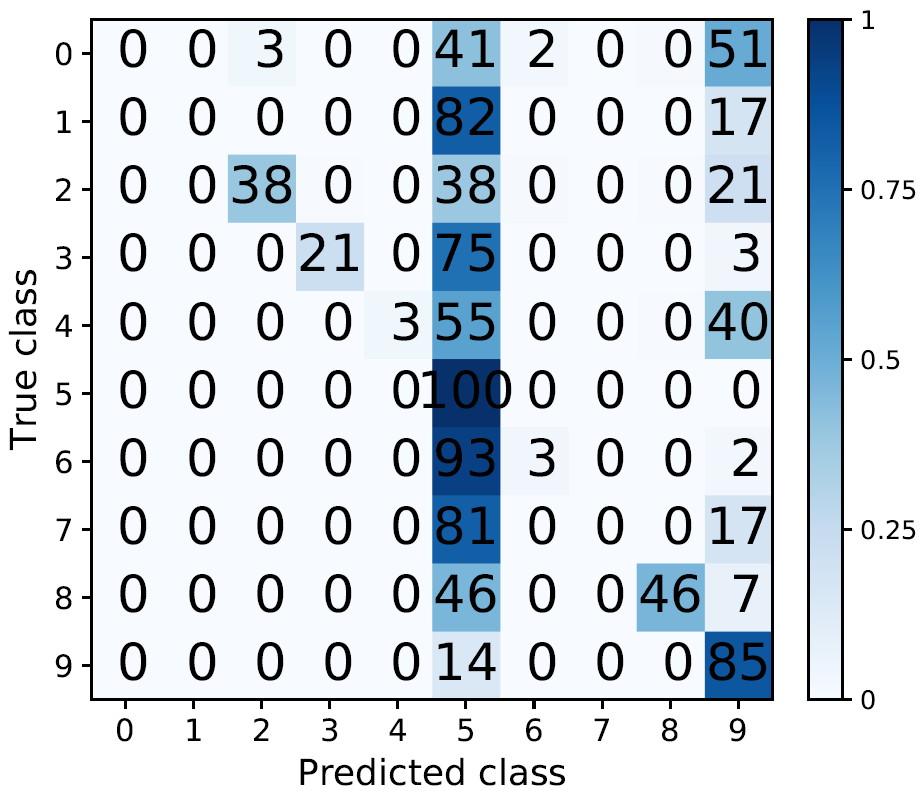}}
\subfigure{\includegraphics[width = 0.46\columnwidth, height = 0.38\columnwidth]{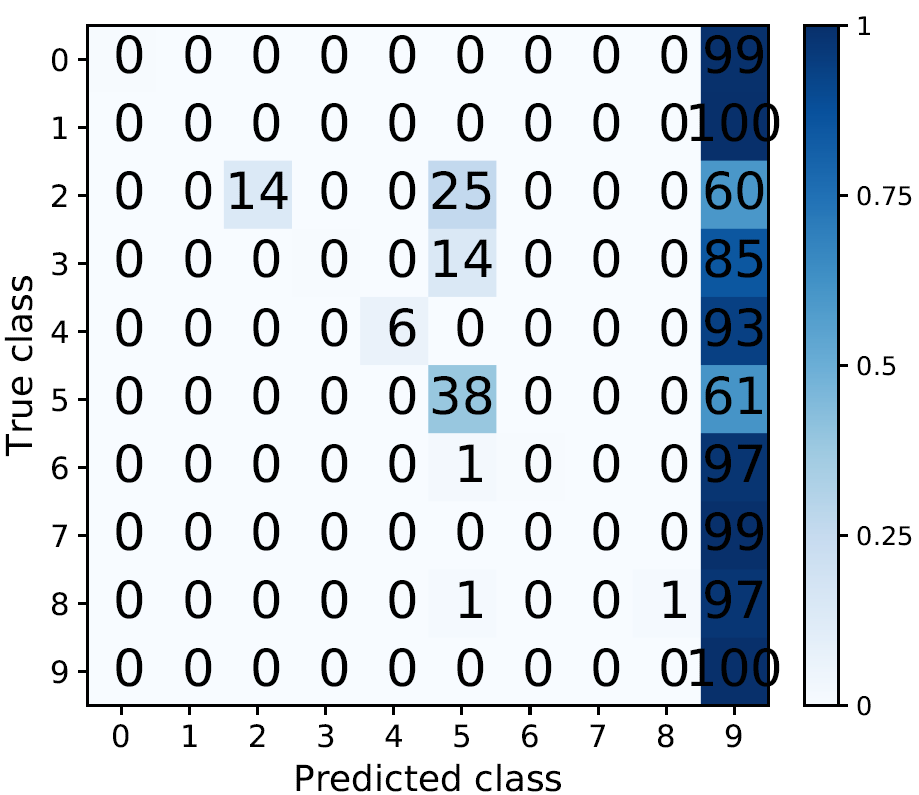}}
\vspace{-0.3cm}
\caption{Accuracy (\%) of the network trained under a two-target backdoor attack  with  $t_1 = 5$ and $t_2 = 9$ ($\alpha = 0.4$ and $\Delta_{tr} = 30$).}
\vspace{-0.3cm}
\label{fig.two_attack}
\end{figure}

We also run some tests by considering a two-target attack. Figure~\ref{fig.two_attack} reports the results we have got when the network is trained under a backdoor attack with target digits $t_1$ = '5' and $t_2$ = 9, corrupted with a ramp and a triangle signal respectively. Both $D_5$ and $D_7$ are attacked with $\alpha = 30$ and $\Delta_{tr} = 30$. In particular, the figure reports the test accuracy when a ramp backdoor (left) and a triangle backdoor (right) is added to the test digits with $\Delta_{ts} = 30$  (we checked that pristine samples are still correctly classified). As we can see, the probability that the networks decides, respectively, for '5' and '9', is rather large. A bit of confusion is made between these two digits when the ramp signal is added (perhaps also due to the similarity of
ramp and triangle signal).
These results are promising, showing that a multiple-target attack is also possible.

\subsection{Traffic Signs classification}%

\begin{figure}
\centering
\subfigure{\includegraphics[width = 0.49\columnwidth]{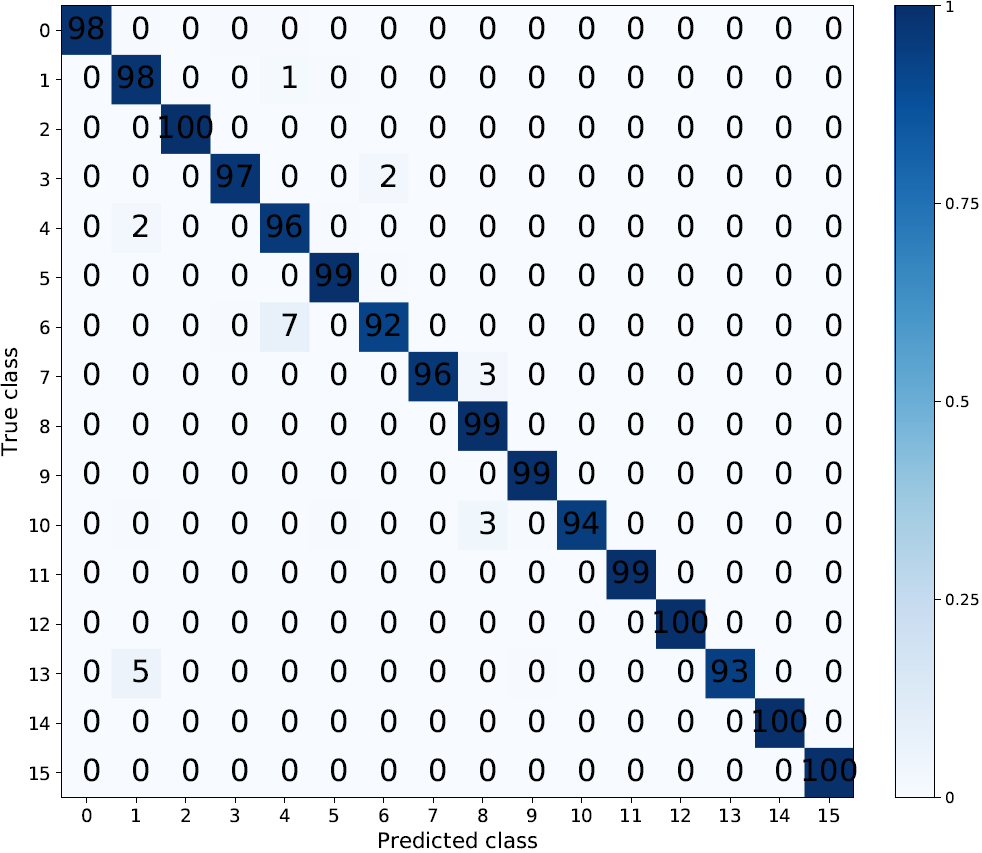}}
\subfigure{\includegraphics[width = 0.49\columnwidth]{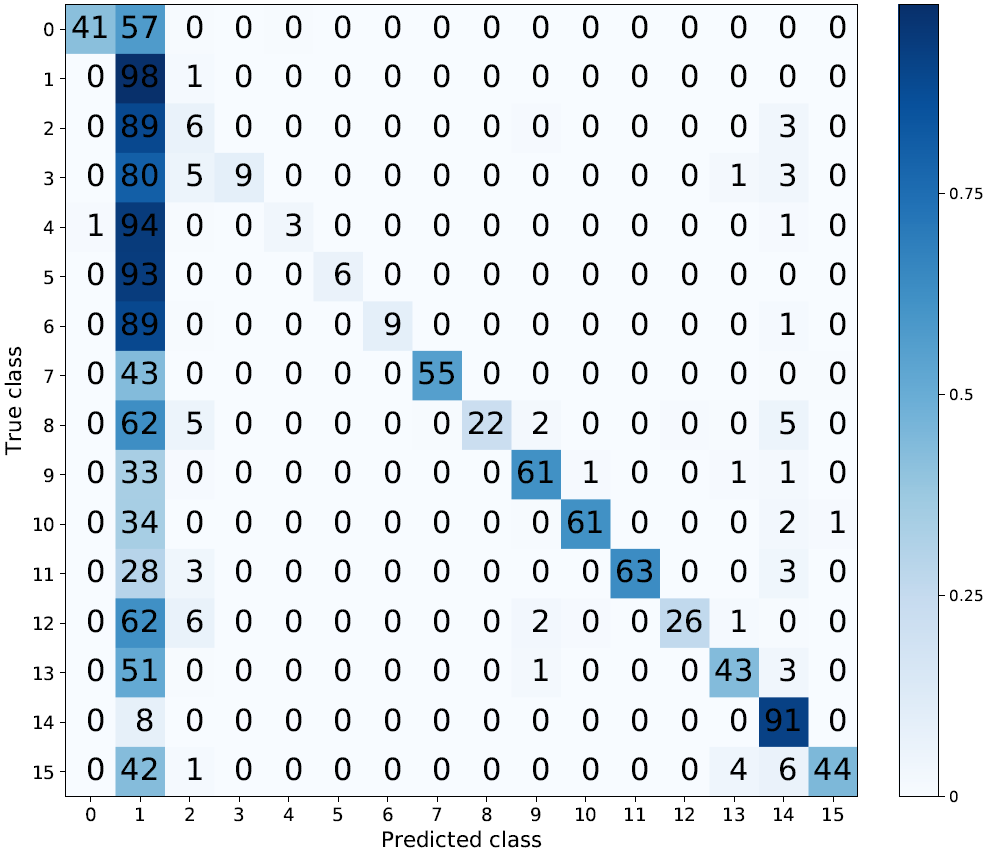}}
\vspace{-0.7cm}
\caption{Accuracy (\%) of the traffic sign classification network trained under a backdoor attack ($\alpha = 0.2$, $\Delta = 20$, $f = 6$), in the absence of attacks at test time (a), in the presence of a backdoor attack with $\Delta_{ts} = 30$ (b).}
\vspace{-0.3cm}
\label{fig.exampleTS}
\end{figure}

Similar tests were carried out for the case of traffic sign classification.
Figure \ref{fig.exampleTS} reports the results we have obtained by letting the target class be the speed limit 50 sign.
The attack has been implemented by letting $\alpha = 0.2$, $\Delta_{tr}=20$ and $f = 6$. Figure \ref{fig.exampleTS} shows the classification accuracy in the absence of attacks at test time  (left), and when the backdoor signal with $\Delta_{ts}=30$ is added to test samples from all the classes (right). We see that, when the sinusoidal backdoor signal is superimposed, the network  classifies several signs from different classes as the speed limit 50 sign with pretty high probability.  Not surprisingly, the speed limit signs (corresponding to label $0$ to $5$ in the figure) are generally easier to attack.
%
The results of more extensive tests are reported in Table \ref{tab:resTS}  where we show the probability that a test sample with a superimposed backdoor signal is classified as the target traffic sign for several strengths of the superimposed signal $\Delta_{ts}$.
The probability is averaged over the  7 most successfully attacked classes (different from $t$). The results are reported for 4 different target signs, corresponding to 2 speed limits ($t=1$ and $3$), 1 prohibition sign ($t=7$) and 1 danger sign ($t = 13$). In each case, the network is trained by letting $\alpha = 0.2$, $\Delta_{tr}=20$. The pristine samples are correctly classified (the accuracy is always larger than 0.95 for every class).
Upon inspection of the table, we see that the network learns the backdoor signal; however, in order to be regarded to as a discriminant feature  (and induce the network to change the decision), in many cases, the backdoor has to be superimposed with a rather large strength at test time. 
We also verified that, by increasing $\alpha$,
the attack performance does not improve much and the classification accuracies remains similar.
Obviously, the effectiveness of the attack can be improved by increasing $\Delta_{tr}$, at the price of a reduced stealthiness.


We also trained the network for traffic signs classification with a two-target backdoor attack.
When the target classes are $t_1 = 1$ and $t_2 = 7$, and a sinusoidal backdoor signal is considered at frequency $f=6$ and $f=3$ respectively (with both $D_1$ and $D_7$ attacked with $\alpha = 0.2$ and $\Delta_{tr} = 20$), the attack success rate (averaged on the 7 best results)  is $80\%$ for $t_1$  and $56\%$ for $t_2$ when $\Delta_{ts}= 40$, $90\%$ for $t_1$ and $67\%$ for $t_2$  when $\Delta_{ts} = 60$.

\begin{table}
\centering
\caption{Attack success rate (\%) in the case of traffic sign classification for several  $\Delta_{ts}$ ($\alpha = 0.2$, $\Delta_{tr} = 20$, $f=6$).
The rate is averaged on the 7 most successfully attacked test signs.}
\vspace{0.2cm}
\renewcommand\arraystretch{1.1}
\setlength{\tabcolsep}{3pt}
\label{tab:resTS}
{\begin{tabular}{|l|c c c c|c c c c|c c c c|c c c c|} 
\hline
\rowcolor{lightgray}
& \multicolumn{4}{c|}{$t = 1$} & \multicolumn{4}{c|}{$t = 3$} & \multicolumn{4}{c|}{$t = 7$}&  \multicolumn{4}{c|}{$t = 13$}\\
\rowcolor{lightgray}
$\Delta_{ts}$  &  20 & 30 & 40 & 60 & 20 & 30 & 40 & 60 & 20 & 30 & 40 & 60 & 20 & 30 & 40 & 60 \\ \hline
\% & 73 & 81 & 79 & 83 & 39 &  62 & 76 & 87 & 52 & 71 & 83 & 93 & 26 & 48 & 60 & 78  \\ \hline
\end{tabular}}
\vspace{-0.3cm}
\end{table}

\section{CONCLUDING REMARKS} 
\label{sec.con}

We have proposed a new backdoor attack which, as opposed to previous works, does not require that the labels of the corrupted samples are poisoned. In this way, the stealthiness of the attack is greatly improved, since the presence of corrupted training samples can not be revealed by detecting the mismatch between the sample content and its label. The flexibility of the attack is also improved, since at training time the attacker needs only to corrupt samples of the target class, while the choice of the source class can be made at test time. The price to pay with respect to attacks with label corruption is that the percentage of samples that must be corrupted is an order of magnitude larger.  We have implemented the new attack by considering two popular classification tasks, namely digit recognition and traffic sign classification.
In the case of  digit recognition task, we were able to successfully attack the classification networks, while keeping the backdoor signal invisible.
For the traffic sign case, the attack is more difficult; however, our results show that the attack without label poisoning is
is effective to some extent with a nearly invisible backdoor. Especially in this case, the choice of a proper backdoor signal is of great importance.
%
Future works will then focus on a better adaptation of the backdoor signal to the classification task and the target class of the attack, with the aim of reducing the strength of the backdoor signal itself and the percentage of corrupted signals required for a successful attack.  It goes without saying that devising proper mechanisms to identify the presence of backdoors into a trained model is of the outmost importance and is going to receive an increasing attention in the next years.



* The list of authors is provided in alphabetic order.

\bibliographystyle{IEEEbib}
\bibliography{ICIP19}

\end{document}